\documentclass[12pt]{article}
\usepackage{amsmath}
\usepackage[dvips]{graphicx}
\usepackage{geometry}

\begin{document}

\author{A. de Souza Dutra\thanks{
dutra@feg.unesp.br} and R. A. C. Correa\thanks{
fis04132@gmail.com} \\
\\
UNESP Univ Estadual Paulista - Campus de Guaratinguet\'{a} - DFQ\\
Av. Dr. Ariberto Pereira Cunha, 333\\
12516-410 Guaratinguet\'{a} SP Brasil}
\title{Traveling solitons in Lorentz-violating systems}
\maketitle

\begin{abstract}
In this work we present a class of traveling solitons in Lorentz-violating
systems. In the case of Lorentz violating scenarios, it is usual to
construct static solitonic configurations. Here it is shown that it is
possible to construct some traveling solitons which, as it should be
expected, can not be mapped into a static configuration by means of Lorentz
boosts due to its explicit breaking. Furthermore, in the model studied, a
complete set of solutions is obtained. The solutions present a critical
behavior controlled by the choose of an arbitrary integration constant.
\end{abstract}

\section{Introduction}

The study of the problem of Lorentz symmetry breaking appeared in the
physics literature motivated by the fact that the superstring theories
suggest that Lorentz symmetry should be violated at higher energies \cite%
{PRD 39 (1989) 683}. Recently, a large amount of works considering the
impact of some kind of Lorentz symmetry breaking have appeared in the
literature. For instance, some years ago, Carrol, Field and Jackiw \cite{PRD
41 (1990) 1231} addressed the problem with CPT (Charge
conjugation-Parity-Time reversal) symmetry violation. On the other hand,
some impact over the standard model due to Lorentz and CPT symmetries were
discussed by Colladay and Kostelecky \cite{PRD 55 (1997) 6760,PRD 58 (1998)
116002}. Other problem analyzed in the literature is the spontaneous
breaking of the four-dimensional Lorentz invariance of the QED \cite{PRD 71
(2005) 044018}. At this point, it is interesting to mention that a
space-time with torsion interacting with a Maxwell field by means of
Chern-Simons-like term was introduced by the authors in Ref. \cite{PRD 71
(2005) 044018}. In this case, it is possible to explain the optical activity
in the synchrotron radiation emitted by cosmological distant radio sources.

Recently motivated by the problem of Lorentz symmetry violating gauge
theories in connection with gravity models, Boldo \textit{et al}. \cite{PLB
689 (2010) 112} have analyzed the graviton excitations and Lorentz-violating
gravity with cosmological constant. It is \ important to remark that a
considerable effort has been done experimentally to observe signs of the
Lorentz and CPT symmetries violation effects. In fact, in a very recent
work, Maccione, Liberati and Sigh \cite{PRL 105 (2010) 021101} have shown
that experimental data on the photon content of ultrahigh-energy cosmic rays
lead to strong constraints over the Lorentz symmetry violations in stringy
space-time foam models. This was done by studying the time delay between $%
\gamma $ rays of different energies from extragalactic sources. Moreover,
Giulia Gubitosi \textit{et al}. \cite{PRD 82 (2010) 024013} have introduced
an important role in the study of Planck-scale modifications to
electrodynamics characterized by a spacelike symmetry-breaking vector. This
year, several studies involving Lorentz violation has appeared in the
literature \cite{PLB 689 (2010) 112}-\cite{PLB 694 (2010) 149}.

Finally, it is important to remark that nonlinear models which have
topological solutions are very interesting and important in many branches of
physics \cite{Book Rajaraman}-\cite{Book Vachaspati}. In a recent work \cite%
{PRD 74 (2006) 085030} it was shown that some nonlinear models in
two-dimensional space-time were two scalar fields interact in the Lorentz
and CPT violating scenarios present static solitonic configurations. This
was done by generalizing a model presented by Barreto and collaborators \cite%
{PRD 73 (2006) 065015}. Finally, in a very recent work, Bazeia \textit{et
al. }\cite{Phys. D 239 (2010) 942}\textit{\ }also\textit{\ }have analyzed
the effects of the Lorentz violation on topological defects generated by two
real scalar fields. In that case, the Lorentz-violating is induced by a
fixed tensor coefficient that couples the two fields. In all of these
examples, the presented solitonic configurations were static. In this work
we are going to show that it is possible to find nontrivial traveling
solutions in this kind of scenario. This is going to be done by taking as
example a generalization of some models recently discussed in the literature
\cite{PRD 74 (2006) 085030}-\cite{Phys. D 239 (2010) 942}. As a consequence,
we present a class of traveling solitons in Lorentz-violating systems as
well as some static configurations. Finally it is shown that the static
configurations are not the limit of the traveling ones. This is done by
using an approach developed to deal with some classes of nonlinear models in
two-dimensional space-time of two interacting scalar fields which were
presented in \cite{PLB 626 (2005) 249}. In this last reference it was shown
that these systems in $1+1$ dimensions, the socalled orbit equation can be
cast in a form of linear first-order differential equation, thus leading to
general solutions of the system. We also show that the solutions present a
critical behavior controlled by the choose of an arbitrary integration
constant.

\section{The model}

Some years ago, it was presented in \cite{PRD 74 (2006) 085030} a two-field
model in 1+1 dimensions where the Lorentz breaking Lagrangian density
generalizes some results in the literature. That Lagrangian density contains
vector functions with dependence on the dynamical scalar fields. Moreover,
the mentioned vector functions are responsible by the Lorentz symmetry
breaking. On the other hand, in reference \cite{Phys. D 239 (2010) 942}, the
effects of the Lorentz violation on topological defects generated by two
real scalar fields was analyzed too, in this last one the Lagrangian density
has a tensor which it is the term which breaks the Lorentz symmetry. Thus,
in this work we construct a generalized two-field model in 1+1 dimensions
which is described by the Lagrangian density%
\begin{eqnarray}
\mathcal{L} &=&\frac{1}{2}\partial _{\mu }\phi \partial ^{\mu }\phi +\frac{1%
}{2}\partial _{\mu }\chi \partial ^{\mu }\chi -G^{\mu }(\phi ,\chi )\partial
_{\mu }\phi -F^{\mu }(\phi ,\chi )\partial _{\mu }\chi +  \notag \\
&&  \notag \\
&&-\gamma k^{\mu \nu }(\partial _{\mu }\phi \partial _{\nu }\phi +\partial
_{\mu }\chi \partial _{\nu }\chi )-pk^{\mu \nu }\partial _{\mu }\phi
\partial _{\nu }\chi -V(\phi ,\chi ),  \label{e1}
\end{eqnarray}

\noindent where $\mu =0,1$, $G^{\mu }(\phi ,\chi )$ and $F^{\mu }(\phi ,\chi
)$ are vector functions, and $V(\phi ,\chi )$ is the potential. Furthermore,
$k^{\mu \nu }$ is a constant tensor, here represented by a $2\times 2$
matrix, where $\alpha _{1}$, $\alpha _{2}$, $\alpha _{3}$ and $\alpha _{4}$
are arbitrary parameters. In fact, a similar process of breaking the Lorentz
symmetry was put forward by Anacleto \textit{et al. }\cite{PLB 694 (2010)
149}  in a recent work, where the tensor $k^{\mu \nu }$ is a $4\times 4$
matrix, in that case the authors studied the problem of the acoustic black
holes from Abelian Higgs model with Lorentz symmetry breaking. Here, as
advertised, the tensor $k^{\mu \nu }$ is written as%
\begin{equation}
k^{\mu \nu }=\left(
\begin{array}{cc}
\alpha _{1} & \alpha _{2} \\
\alpha _{3} & \alpha _{4}%
\end{array}%
\right) .  \label{e2}
\end{equation}

Note that, from the Lagrangian density (\ref{e1}), we can recover the one
presented in the work by Bazeia \textit{et al} \cite{Phys. D 239 (2010) 942}
by choosing $\gamma =0$, $G^{0}(\phi ,\chi )=F^{0}(\phi ,\chi )=0$, $\alpha
_{1}=\alpha _{4}=\beta $, $\alpha _{2}=\alpha _{3}=\alpha $ and $\ p=-1$.
Furthermore, we can also recover the lagrangian density presented in \cite%
{PRD 74 (2006) 085030, PRD 73 (2006) 065015} by setting conveniently the
above defined parameters. Therefore, we have a more general model including
vector functions and a constant tensor. It is important to remark that the
more general model presented here, can be used to bring more information
about the impact of the Lorentz violation of important systems like, for
instance, those presenting topological structures \cite{PRD 74 (2006)
085030, PRD 73 (2006) 065015}.

From the Lagrangian density (\ref{e1}), we can to write the corresponding
equations of motion
\begin{eqnarray}
&&\left. (1-\gamma ~\alpha _{1})\ddot{\phi}-(1+\gamma ~\alpha _{4})\phi
^{\prime \prime }-p(\alpha _{1}\ddot{\chi}+\alpha _{4}\chi ^{\prime \prime
})+(F_{\phi }^{0}-G_{\chi }^{0})\dot{\chi}+(F_{\phi }^{1}-G_{\chi }^{1})\chi
^{^{\prime }}+\right.  \label{e3} \\
&&  \notag \\
&&\left. -(\alpha _{3}+\alpha _{2})(\gamma \dot{\phi}^{^{\prime }}+p\dot{\chi%
}^{^{\prime }})+V_{\phi }=0,\right.  \notag \\
&&  \notag \\
&&\left. (1-\gamma ~\alpha _{1})\ddot{\chi}-(1+\gamma ~\alpha _{4})\chi
^{\prime \prime }-p(\alpha _{1}\ddot{\phi}+\alpha _{4}\phi ^{\prime \prime
})-(F_{\phi }^{0}-G_{\chi }^{0})\dot{\phi}-(F_{\phi }^{1}-G_{\chi }^{1})\phi
^{^{\prime }}+\right.  \label{e4} \\
&&  \notag \\
&&\left. -(\alpha _{3}+\alpha _{2})(\gamma ~\dot{\chi}^{^{\prime }}+p~\dot{%
\phi}^{^{\prime }})+V_{\chi }=0,\right.  \notag
\end{eqnarray}

\noindent where the dot stands for derivative with respect to time, while
the prime represents derivative with respect to $x$, $V_{\phi }\equiv
\partial V/\partial \phi $ and $V_{\chi }\equiv \partial V/\partial \chi $.
It can be seen that the two equations in above are carrying informations of
the Lorentz breaking of the model through the presence of the $\alpha _{i}$
parameters and the vector functions. But, as a consequence of the model
studied in this work, in general we can not solve analytically the above
differential equations. However one can consider an interesting case for the
fields configurations, where one searches for traveling waves solutions.
Configurations that exhibit traveling waves has an important impact when we
study boundary states for D-branes and the supergravity fields for a D-brane
\cite{JHEP 04 (2003) 032}-\cite{PLB 694 (2010) 246}.

Then, let us begin our search for traveling waves solutions in the form $%
\phi =\phi (u)$ and $\chi =\chi (u)$ with $u=Ax+Bt$. Thus, the equations (%
\ref{e3}) and (\ref{e4}) take the form%
\begin{eqnarray}
-\phi _{uu}+\tilde{\beta}\chi _{uu}-\tilde{\alpha}\chi _{u}+\tilde{V}_{\phi
} &=&0,  \label{e5} \\
&&  \notag \\
-\chi _{uu}+\tilde{\beta}\phi _{uu}+\tilde{\alpha}\phi _{u}+\tilde{V}_{\chi
} &=&0,  \label{e6}
\end{eqnarray}

\noindent with the definitions%
\begin{eqnarray}
\tilde{\beta} &\equiv &-\frac{p[(\alpha _{2}+\alpha _{3})AB+\alpha
_{4}A^{2}+\alpha _{1}B^{2}]}{(1+\gamma \alpha _{4})A^{2}-(1-\gamma \alpha
_{1})B^{2}+AB\gamma (\alpha _{2}+\alpha _{4})},\text{ \ }  \label{e7} \\
&&  \notag \\
\tilde{\alpha} &\equiv &-\frac{B(F_{\phi }^{0}-G_{\chi }^{0})+A(F_{\phi
}^{1}-G_{\chi }^{1})}{(1+\gamma \alpha _{4})A^{2}-(1-\gamma \alpha
_{1})B^{2}+AB\gamma (\alpha _{2}+\alpha _{4})}, \\
&&  \notag \\
\tilde{V}_{\phi } &\equiv &\frac{V_{\phi }}{(1+\gamma \alpha
_{4})A^{2}-(1-\gamma \alpha _{1})B^{2}+AB\gamma (\alpha _{2}+\alpha _{4})},%
\text{ \ \ } \\
&&  \notag \\
\tilde{V}_{\chi } &\equiv &\frac{V_{\chi }}{(1+\gamma \alpha
_{4})A^{2}-(1-\gamma \alpha _{1})B^{2}+AB\gamma (\alpha _{2}+\alpha _{4})}.
\end{eqnarray}

In order to decouple the pair of second order differential equations, we
multiply the equation (\ref{e5}) by $\phi _{u}$ and the equation (\ref{e6})
by $\chi _{u}$. Thus, it is not difficult to conclude that, after adding the
two equations, one can write%
\begin{equation}
\frac{d}{du}\left[ -\frac{1}{2}(\phi _{u}^{2}+\chi _{u}^{2})+\tilde{\beta}%
\phi _{u}\chi _{u}+\tilde{V}(\phi ,\chi )\right] =0.  \label{e8}
\end{equation}

In this case, we have%
\begin{equation}
-\frac{1}{2}(\phi _{u}^{2}+\chi _{u}^{2})+\tilde{\beta}\phi _{u}\chi _{u}+%
\tilde{V}(\phi ,\chi )=c_{0}.  \label{e9}
\end{equation}

The above equation can be rewritten for the case where $c_{0}=0$ which is
necessary in order to allow solitonic solutions, otherwise one obtains
oscillating or complex solutions \cite{PRL 100 (2008) 041602}. Therefore, we
get%
\begin{equation}
-\frac{1}{2}(\phi _{u}^{2}+\chi _{u}^{2})+\tilde{\beta}\phi _{u}\chi _{u}+%
\tilde{V}(\phi ,\chi )=0.  \label{e10}
\end{equation}

Note that in the above equation, the dependence in $\tilde{\alpha}$ has
disappeared. However, the dependence of the system in terms of the Lorentz
breaking parameters is still present but it is implicit. Now, in order to
desacouple the above equation, we apply the rotation%
\begin{equation}
\left(
\begin{array}{c}
\phi (u) \\
\chi (u)%
\end{array}%
\right) =\frac{1}{\sqrt{2}}\left(
\begin{array}{cc}
1 & -1 \\
1 & 1%
\end{array}%
\right) \left(
\begin{array}{c}
\theta (u) \\
\varphi (u)%
\end{array}%
\right) ,  \label{e11}
\end{equation}

\noindent thus, the equation (\ref{e10}) is rewritten as%
\begin{equation}
-\frac{1}{2}(1-\tilde{\beta})\theta _{u}^{2}-\frac{1}{2}(1+\tilde{\beta}%
)\varphi _{u}^{2}+\tilde{V}(\theta ,\varphi )=0.  \label{e12}
\end{equation}

Furthermore, performing the dilations%
\begin{equation}
\theta (u)=\frac{\sigma (u)}{\sqrt{1-\tilde{\beta}}},\text{ \ \ \ \ }\varphi
(u)=\frac{\rho (u)}{\sqrt{1+\tilde{\beta}}},  \label{e13}
\end{equation}

\noindent one gets%
\begin{equation}
-\frac{1}{2}\sigma _{u}^{2}-\frac{1}{2}\rho _{u}^{2}+\tilde{V}(\sigma ,\rho
)=0.  \label{e14}
\end{equation}

At this point one can verify that the above equation allow one to write two
first-order coupled differential equations. In this case it is usual to
impose that the potential must be written in terms of a superpotential like%
\begin{equation}
\bar{V}(\sigma ,\rho )=\frac{1}{2}\left( \frac{\partial W(\sigma ,\rho )}{%
\partial \sigma }\right) ^{2}+\frac{1}{2}\left( \frac{\partial W(\sigma
,\rho )}{\partial \rho }\right) ^{2},
\end{equation}%
which leads to the following set of equations%
\begin{equation}
\frac{d\sigma }{du}=\pm W_{\sigma },\text{ \ \ \ }\frac{d\rho }{du}=\pm
W_{\rho },  \label{e19}
\end{equation}

\noindent where $W_{\sigma }\equiv \partial W(\sigma ,\rho )/\partial \sigma
$ and $W_{\rho }\equiv \partial W(\sigma ,\rho )/\partial \rho $, and this
will leads us to the solitonic solutions we are looking for.

In order to analyze the energy of the configurations obtained, we write the
energy-momentum tensor in the form%
\begin{equation}
T^{\mu \nu }=\frac{\partial \mathcal{L}}{\partial (\partial _{\mu }\phi )}%
\partial ^{\nu }\phi +\frac{\partial \mathcal{L}}{\partial (\partial _{\mu
}\chi )}\partial ^{\nu }\chi -g^{\mu \nu }\mathcal{L}\text{.}
\end{equation}

Therefore, the energy density for the lagrangian density (\ref{e1}) is given
by
\begin{eqnarray}
T^{00} &=&\frac{\dot{\phi}}{2}^{2}+\frac{\dot{\chi}^{2}}{2}+\left( \frac{1}{2%
}+\gamma \alpha _{4}\right) \left( \phi ^{^{\prime }2}+\chi ^{^{\prime
}2}\right) +G^{1}(\phi ,\chi )\phi ^{^{\prime }}+F^{1}(\phi ,\chi )\chi
^{^{\prime }}+  \notag \\
&&  \notag \\
&&-p\alpha _{1}\dot{\phi}\dot{\chi}+\gamma (\alpha _{2}+\alpha _{3})\phi
^{^{\prime }}\dot{\phi}+\gamma (\alpha _{2}+\alpha _{3})\chi ^{^{\prime }}%
\dot{\chi}+p\alpha _{4}\phi ^{^{\prime }}\chi ^{^{\prime }}+p\alpha _{2}\dot{%
\phi}\chi ^{^{\prime }}+  \notag \\
&&  \notag \\
&&+p\alpha _{3}\phi ^{^{\prime }}\dot{\chi}+V(\phi ,\chi ).
\end{eqnarray}

For the traveling waves solutions, the energy density is written in the form%
\begin{eqnarray}
&&\left. T_{traveling}^{00}=\left[ \frac{B^{2}}{2}+\left( \frac{1}{2}+\gamma
\alpha _{4}\right) A^{2}+\gamma (\alpha _{2}+\alpha _{3})AB\right] \phi
_{u}^{2}\right.  \notag \\
&&  \notag \\
&&\left. +\left[ \frac{B^{2}}{2}+\left( \frac{1}{2}+\gamma \alpha
_{4}\right) A^{2}+\gamma (\alpha _{2}+\alpha _{3})AB\right] \chi _{u}^{2}+p%
\left[ \alpha _{4}A^{2}-\alpha _{1}B^{2}+(\alpha _{2}+\alpha _{3})AB\right]
\phi _{u}\chi _{u}\right.  \notag \\
&&  \notag \\
&&\left. +A[G^{1}(\phi ,\chi )\phi _{u}+F^{1}(\phi ,\chi )\chi _{u}]+V(\phi
,\chi ).\right.
\end{eqnarray}

Now, we choose the superpotential which was used in \cite{PLB 626 (2005) 249}%
, which is written as%
\begin{equation}
W(\sigma ,\rho )=-\lambda \sigma +\frac{\lambda }{3}\sigma ^{3}+\mu \sigma
\rho ^{2}.  \label{e20}
\end{equation}

In this case, the solutions presented in Ref. \cite{PLB 626 (2005) 249},
with $\lambda =\mu $, are given by%
\begin{eqnarray}
\sigma _{+}(u) &=&\frac{(c_{0}^{2}-4)e^{4\mu (u-u_{0})}-1}{[c_{0}e^{2\mu
(u-u_{0})}-1]^{2}-4e^{4\mu (u-u_{0})}},\sigma _{-}(u)=\frac{%
4-c_{0}^{2}+e^{4\mu (u-u_{0})}}{[e^{2\mu (u-u_{0})}-c_{0}]^{2}-4},  \notag \\
&&  \label{e21} \\
\rho _{+}(u) &=&\frac{4e^{2\mu (u-u_{0})}}{[c_{0}e^{2\mu
(u-u_{0})}-1]^{2}-4e^{4\mu (u-u_{0})}},\rho _{-}(u)=\frac{4e^{2\mu (u-u_{0})}%
}{[e^{2\mu (u-u_{0})}-c_{0}]^{2}-4},  \notag
\end{eqnarray}

\noindent where we must impose that $c_{0}\leq -2$ in both solutions. On the
other hand, in the case where $\lambda =4\mu $, the exact solutions are
written as%
\begin{eqnarray}
\sigma _{+}(u) &=&\frac{4+(16c_{0}-1)e^{8\mu (u-u_{0})}}{[2+e^{4\mu
(u-u_{0})}]^{2}-16c_{0}e^{8\mu (u-u_{0})}},\sigma _{-}(u)=\frac{%
16c_{0}+4e^{8\mu (u-u_{0})}-1}{[1+2e^{4\mu (u-u_{0})}]^{2}-16c_{0}},  \notag
\\
&&  \label{e23} \\
\rho _{+}(u) &=&-\frac{2e^{2\mu (u-u_{0})}}{\sqrt{[(1/2)e^{4\mu
(u-u_{0})}+1]^{2}-4c_{0}e^{8\mu (u-u_{0})}}},\rho _{-}(u)=-\frac{4e^{2\mu
(u-u_{0})}}{\sqrt{[1+2e^{4\mu (u-u_{0})}]^{2}-16c_{0}}}.  \notag
\end{eqnarray}

\noindent In this case, we impose that $c_{0}\leq 1/16$. It is important to
remark that making the exchange of $\sigma \rightarrow \rho $ and $\rho
\rightarrow \sigma $ in the case where $\lambda =\mu $, the equation of
motion (\ref{e14}) remains invariant. Thus, the solutions were kinks become
lumps and \textit{vice-versa} shall appear, and this is used in order to
generate the complete set of orbits appearing in the Figure 4. In fact, this
symmetry is important for the generation of the complete set of orbits
connecting the vacua.

Thus, the fields $\phi (u)$ and $\chi (u)$ are given by%
\begin{eqnarray}
\phi _{\pm }(u) &=&\frac{1}{\sqrt{2}}\left[ \frac{\sigma _{\pm }(u)}{\sqrt{1-%
\tilde{\beta}}}-\frac{\rho _{\pm }(u)}{\sqrt{1+\tilde{\beta}}}\right] ,
\notag \\
&&  \label{e24} \\
\chi _{\pm }(u) &=&\frac{1}{\sqrt{2}}\left[ \frac{\sigma _{\pm }(u)}{\sqrt{1-%
\tilde{\beta}}}+\frac{\rho _{\pm }(u)}{\sqrt{1+\tilde{\beta}}}\right] .
\notag
\end{eqnarray}

Now, using the solutions presented in the reference \cite{PLB 626 (2005) 249}
which are represented here by (\ref{e21}) and (\ref{e23}), we have the
complete set of solutions with position and time dependence.

Here, we call attention to the fact that the static solutions for the
equations (\ref{e3}) and (\ref{e4}) are different from the traveling wave
ones. This difference can be seen from an inspection of the static fields
differential equations%
\begin{eqnarray}
-\phi ^{\prime \prime }+\bar{\beta}\chi ^{\prime \prime }+\bar{\alpha}\chi
^{^{\prime }}+\bar{V}_{\phi } &=&0,  \label{e25} \\
&&  \notag \\
-\chi ^{\prime \prime }+\bar{\beta}\phi ^{\prime \prime }-\bar{\alpha}\chi
^{^{\prime }}+\bar{V}_{\chi } &=&0,  \label{e26}
\end{eqnarray}

\noindent where now, one have%
\begin{eqnarray}
\bar{\beta} &\equiv &\frac{-p\alpha _{4}}{(1+\gamma \alpha _{4})},\bar{\alpha%
}\equiv \frac{(F_{\phi }^{1}-G_{\chi }^{1})}{(1+\gamma \alpha _{4})},  \notag
\\
&& \\
\bar{V}_{\phi } &\equiv &\frac{V_{\phi }}{(1+\gamma \alpha _{4})}\text{ \ \
\ \ and \ \ \ \ \ \ \ }\bar{V}_{\chi }\equiv \frac{V_{\chi }}{(1+\gamma
\alpha _{4})}.  \notag
\end{eqnarray}

In particular, if $\gamma =0$, $G^{\mu }(\phi ,\chi )=F^{\mu }(\phi ,\chi
)=0 $, $\alpha _{1}=\alpha _{4}=\beta $, $\alpha _{2}=\alpha _{3}=\alpha $
and $p=-1$, ones recovers the model presented by \ Bazeia \textit{et al}
\cite{Phys. D 239 (2010) 942}\textit{. }In the static case, the equations of
motion presented by the authors are given by%
\begin{eqnarray}
-\phi ^{\prime \prime }+\beta \chi ^{\prime \prime }+V_{\phi } &=&0,
\label{e27} \\
&&  \notag \\
-\chi ^{\prime \prime }+\beta \phi ^{\prime \prime ^{\prime }}+V_{\chi }
&=&0.  \label{e28}
\end{eqnarray}

Here, it is \ interesting to note that the pair of equations presented in
\cite{Phys. D 239 (2010) 942} for the static solutions takes on a different
form compared with the equations (\ref{e25}) and (\ref{e26}). In fact, we
can recover the equations of motion presented in the work of Bazeia \cite%
{Phys. D 239 (2010) 942} by setting the correct parameters. But the general
static configurations are given by equations (\ref{e25}) and (\ref{e26}),
which are carrying more informations of the terms of the Lorentz breaking of
the model.

The difference discussed above can be seen in Figures 1 and 2. In Figure 3
the complete set of classes of orbits are illustrated.

\section{Conclusions}

In this work we have shown that a class of traveling solitons in
Lorentz-violating systems can be analytically obtained, which happens
despite the fact that there is no Lorentz symmetry and consequently one can
not recover the traveling solutions from the static one, just performing
Lorentz boosts. This has been done by using nonlinear models in
two-dimensional space-time of two interacting scalar fields which were
presented in \cite{PLB 626 (2005) 249}. Furthermore, in the model studied, a
complete set of solutions was obtained. The solutions present a critical
behavior controlled by the choose of an arbitrary integration constant.

\bigskip

\textbf{Acknowledgments:} The authors thanks to CNPq and CAPES for partial
financial support.

{\large \newpage }

\begin{figure*}[tbp]
{\large \centering
\includegraphics [width=16.0cm]{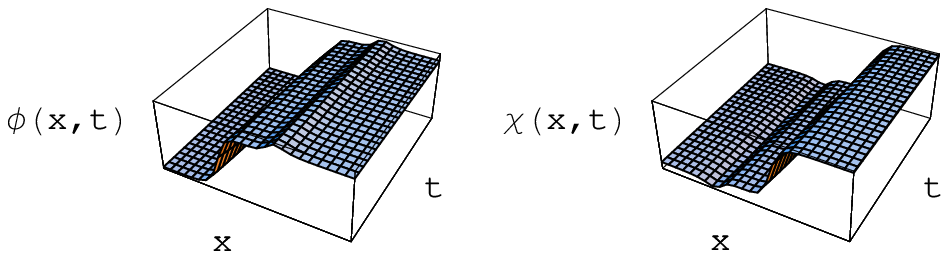} }
\caption{Solution $\protect\phi(x,t)$ and $\protect\chi(x,t)$, with position
and time dependence for $\protect\lambda=4\protect\mu$, $\protect\mu=0.5$, $%
c_{0}=1/16.0001$, $A=0.5$, $B=-0.1$ and $\protect\beta=0.7$.}
\label{Figura2:}
\end{figure*}

\begin{figure*}[tbp]
{\large \centering
\includegraphics[width=18.0cm]{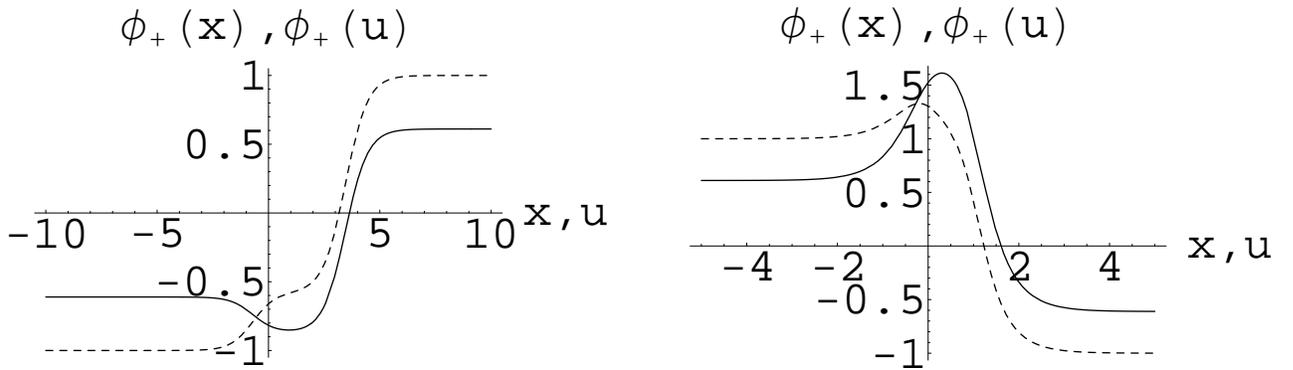} }
\caption{Traveling solitons solutions and static solutions for $\protect%
\lambda=\protect\mu=1$ and $c_{0}=-2.001$. The dashed line corresponds to
the static case with $\protect\beta=0.5$, $A=1$. The thin continuous line
corresponds to the traveling wave case for $\protect\beta=0.5$, $\protect%
\alpha=0.4$, $A=1$, $B=-1.5$.}
\label{Figura3:}
\end{figure*}

\begin{figure*}[tbp]
{\large \centering
\includegraphics {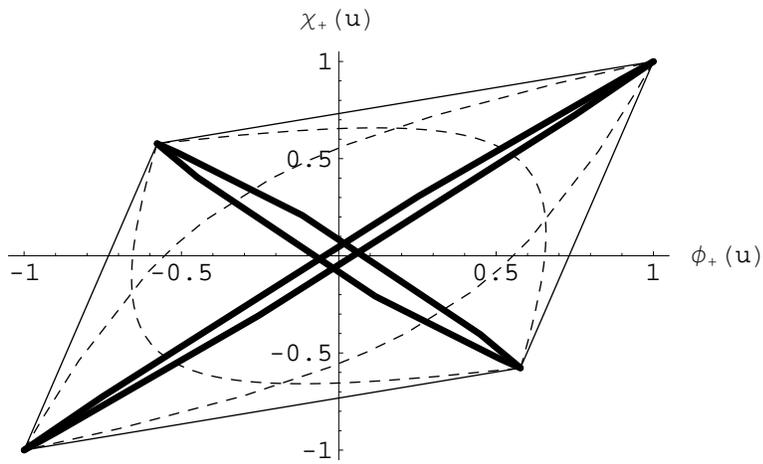} }
\caption{Orbit for the solutions with $\protect\lambda =\protect\mu =1$ and $%
\tilde{\protect\beta}=0.5$. The thin line corresponds to the case where $%
c_{0}=-2.00001$, the dash line corresponds to the case where $%
c_{0}=-2.4$ and the thick line with $c_{0} =-20$.}\label{Figura4:}
\end{figure*}

\end{document}